%% file: main.tex
\documentclass[11pt]{article}
\usepackage{macros}

\begin{document}

\title{Eavesdropper-Blind Remote State Preparation \\ and Applications to Quantum Public-Key Encryption}
\author{Kaniuar Bacho\thanks{University of Edinburgh. Email: k.bacho@sms.ed.ac.uk} \qquad
Alexandru Cojocaru\thanks{University of Edinburgh. Email: a.cojocaru@ed.ac.uk}
}
\date{}
\maketitle
\begin{abstract}
    \input{sec-abstract}

\end{abstract}

\clearpage
\tableofcontents

\clearpage

\input{sec-intro}
\input{sec-prelim}
\input{sec-rsp}
\input{sec-qpke}

\section*{Acknowledgments}
The authors thank Tomoyuki Morimae for several helpful discussions at many different stages of the project and valuable feedback on drafts of this paper, the anonymous ASIACRYPT 2026 reviewers for their valuable comments and suggestions, and Giulio Malavolta for helpful discussions. The authors acknowledge support from the National Science Foundation grant CCF-1813814, from the AFOSR under Award Number FA9550-20-1-0108 and from the Quantum Advantage Pathfinder project.

\clearpage
\bibliographystyle{alpha}
\bibliography{abbrev2,crypto,ref}

\clearpage
\appendix
\input{sec-appendix}

\end{document}

%% file: sec-abstract.tex
Remote state preparation (RSP) is a central primitive in quantum cryptography, enabling classical parties to remotely construct quantum states using only classical communication. As a result, RSP serves as a key building block in numerous protocols involving classical clients and quantum servers, allowing classical parties to leverage the advantages offered by powerful quantum computers. All known constructions of RSP rely on strong cryptographic assumptions, typically variants of trapdoor claw-free functions (TCFs).

In this work, we initiate the study of a weaker form of remote state preparation, which we call \emph{eavesdropper-blind remote state preparation} (EB-RSP). Informally, EB-RSP requires blindness only against external observers who see the transcript of the honest protocol, rather than against the quantum server itself. Despite this relaxed adversarial model, the resulting notion remains sufficient for useful cryptographic applications. In particular, we show that two-message EB-RSP already suffices to construct quantum public-key encryption with classical public keys and quantum ciphertexts. We then construct two-message EB-RSP protocols from specific one-way group actions, yielding a first step toward RSP-type primitives based on assumptions that do not rely on trapdoors. Finally, we observe that existing RSP constructions are likely naturally adaptable to the two-message EB-RSP notion; we demonstrate this explicitly for a concrete TCF-based RSP construction.

%% file: sec-intro.tex
\section{Introduction}\label{sec:intro}
\emph{Remote state preparation} (RSP) is a central quantum cryptographic primitive that enables a classical party to remotely prepare quantum states on a quantum device using only classical communication. As a result, RSP plays a fundamental role in the study of the \emph{Quantum-Computation Classical-Communication} (QCCC) model \cite{C:ACCFLM22}, whose goal is to enable classical parties to leverage the advantages provided by powerful quantum computers. RSP serves as a key building block in numerous protocols in the QCCC setting, including delegated and verifiable quantum computation \cite{FOCS:GheVid19,C:BarKhu25,C:BKMSW25,ITCS:Zhang25}, proofs of quantumness \cite{EPRINT:AMMW22,C:BarKhu25}, quantum fully homomorphic encryption \cite{C:GupVai24}, quantum public-key encryption \cite{C:BarKhu25}, and unclonable cryptography \cite{ICALP:GheMetPor23}.

An RSP protocol \cite{AC:CCKW19,FOCS:GheVid19,CCKP21} is an interactive protocol between a classical verifier (or client) and a quantum prover (or server). At the end of the protocol, the prover should hold a quantum state from a prescribed family of states, while the verifier obtains a classical description of the output quantum state. Existing RSP constructions differ along two main axes:
\begin{enumerate}
    \item The family of quantum states that can be remotely prepared, and
    
    \item The security guarantees provided by the protocol.
\end{enumerate}
These two aspects largely determine the types of applications for which a given RSP protocol is suitable. Most existing constructions prepare BB84 states \cite{C:BarKhu25}, generalized BB84 states \cite{AC:CCKW19,FOCS:GheVid19,EPRINT:AMMW22,C:BKMSW25,ITCS:Zhang25}, or EPR pairs \cite{C:GupVai24}. On the security side, two central notions are \emph{blindness} and \emph{verifiability}. Informally, blindness requires that the prover learns essentially no information about the prepared quantum state, while verifiability guarantees that a successful prover has prepared a valid state consistent with the verifier's classical description.

There are also several possible threat models for blindness. One of the weakest notions is \emph{honest-but-curious blindness} (or \emph{semi-honest blindness}), where the prover follows the protocol honestly but later attempts to infer information about the hidden state. Stronger notions consider \emph{malicious-server blindness}, where the prover may arbitrarily deviate from the protocol and still should not learn information about the prepared state. Existing RSP constructions typically aim to achieve such strong malicious-server security notions.

However, different applications may require different levels of security. This motivates the investigation of weaker forms of remote state preparation that may still suffice for meaningful cryptographic applications while potentially admitting constructions from weaker assumptions.

A long line of work has sought to understand which cryptographic assumptions suffice to realize remote state preparation. Existing constructions typically rely on trapdoor cryptographic primitives, notably trapdoor claw-free functions (TCFs) or variations thereof, while reducing the required assumptions has remained a major open problem. On the other hand, recent impossibility results suggest that one-way functions alone are unlikely to suffice for constructing certain families of RSP protocols \cite{C:BarKhu25}.

This naturally raises the question of whether weaker forms of remote state preparation may still suffice for meaningful cryptographic applications while admitting constructions from weaker assumptions. More concretely, two closely related questions emerge:
\begin{quote}
    (1) \emph{Can useful cryptographic applications still be derived from RSP protocols that satisfy weaker security notions?}
\end{quote}
\begin{quote}
    (2) \emph{Can such weaker forms of RSP be constructed from trapdoor-free cryptographic assumptions?}
\end{quote}

This work investigates both questions. We first introduce a weaker notion of remote state preparation, called \emph{eavesdropper-blind remote state preparation} (EB-RSP), and show that despite its relaxed security guarantees, it still suffices for meaningful cryptographic applications. We then show that such protocols can be constructed from \emph{specific} one-way group actions\footnote{The \emph{specific} one-way group action can be found in \cref{thm:owga-rsp}.}, yielding a first step toward RSP-type primitives that do not rely on trapdoor assumptions.

The importance of the first question lies in understanding whether meaningful cryptographic applications can already be obtained from weaker forms of remote state preparation. Such weaker notions may potentially admit constructions based on qualitatively different or weaker cryptographic assumptions.

Our first contribution addresses the question of whether weaker notions of remote state preparation can still support meaningful cryptographic applications. We introduce a new primitive called \emph{eavesdropper-blind remote state preparation} (EB-RSP), based on what we argue is one of the weakest natural security notions for remote state preparation: blindness only against an external eavesdropper observing the honest transcript of the interaction between the verifier and the prover. Concretely, the security guarantee ensures that, given the transcript of the protocol, any quantum polynomial-time eavesdropper cannot infer any information about the classical description of the prepared quantum state. Unlike standard blindness notions for RSP, EB-RSP does not require hiding the prepared state from the prover itself. A central message of this work is that such a relaxed adversarial model may nevertheless suffice for useful cryptographic applications.

In particular, we show that any two-message EB-RSP protocol implies quantum public-key encryption (QPKE) with classical public keys and quantum ciphertexts. In the classical setting, \cite{STOC:ImpRud89} showed that classical public-key encryption cannot be constructed from one-way functions in a black-box manner. However, when allowing quantum public keys and quantum communication, it is known that one-way functions suffice to construct quantum public-key encryption \cite{TCC:BGHMSV23,EPRINT:BarMalWal23,C:KMNY24}. Nevertheless, quantum public keys introduce substantial challenges compared to classical public-key infrastructures, where public keys can be freely copied, stored, and distributed through classical communication channels.

These limitations motivate the study of QPKE with classical public keys. Previous work \cite{EC:HhaMorYam23} constructed such schemes from non-abelian pseudorandom group actions. In contrast, our work identifies two-message EB-RSP as an intermediate primitive that suffices for constructing QPKE in a black-box manner. As we show later, two-message EB-RSP itself can be instantiated from specific one-way group actions and can also be obtained by adapting existing TCF-based RSP protocols. By combining these ingredients, we obtain constructions of QPKE from certain one-way group actions and from TCFs.

Our second main contribution addresses the possibility of constructing RSP-type primitives without relying on trapdoor cryptographic assumptions. From a foundational perspective, this question is particularly compelling, as all existing RSP constructions rely on cryptographic primitives that admit a trapdoor in order to allow the classical verifier to efficiently recover a classical description of the prepared quantum state. In contrast, we show that EB-RSP protocols can be constructed from specific one-way group actions.

At a conceptual level, our proposed construction replaces the role traditionally played by the trapdoor with algebraic structure arising from the group action. More specifically, we exploit the algebraic properties of the group action to eliminate parameters unknown to the verifier from the description of the output quantum state, while still preserving security. To the best of our knowledge, this is the first construction of an RSP-type primitive with meaningful cryptographic applications that does not rely on any trapdoor primitive.

More broadly, our results suggest that algebraic cryptographic assumptions may provide an alternative route toward quantum cryptographic primitives in the QCCC model. While existing constructions of RSP rely heavily on trapdoor claw-free functions, one-way group actions arise from a fundamentally different framework  rooted in group-based and isogeny-based cryptography.

Group actions have long served as a versatile abstraction in cryptography, starting with the seminal work of Brassard and Yung \cite{C:BraYun90}, by generalizing Diffie--Hellman-type constructions \cite{MazeMonicoRosenthal2007} and playing a central role in isogeny-based cryptography \cite{PQCRYPTO:JaoDeF11,AC:CLMPR18,AC:ADMP20}. Group-action-based assumptions have also found applications in quantum cryptography. For instance, stronger assumptions such as pseudorandom group actions have been used to construct QPKE with classical public keys \cite{EC:HhaMorYam23}, while related assumptions have enabled constructions of quantum money and quantum lightning schemes \cite{AC:MonSha24,ITCS:Zhandry24a,STOC:BosNehZha25}. More recently, several works introduced \emph{quantum group actions} \cite{EPRINT:MorXag24,C:MutZha25}, extending group actions to quantum settings and enabling additional applications, including pseudorandom quantum states and quantum key distribution.

Our work therefore provides evidence that algebraic techniques based on group actions may be useful for constructing QCCC cryptographic primitives beyond the trapdoor-based paradigm. A naturally arising open question is whether stronger forms of remote state preparation, such as standard blind or verifiable RSP, can also be constructed from trapdoor-free assumptions.

\subsection{Main Results and Technical Overview}\label{subsec:technical-outline}
In this section, we describe our main contributions and the key ideas underlying these results. We first introduce a new family of remote state preparation protocols, called \emph{eavesdropper-blind remote state preparation} (EB-RSP), and show how to construct a \emph{two-message} version of EB-RSP from certain one-way group actions (OWGAs) and, as a side result, also from plain trapdoor claw-free functions (TCFs). We then explain the role of EB-RSP and, in particular, show that the two-message version is sufficient for constructing quantum public-key encryption (QPKE) schemes with classical public keys and quantum ciphertexts. By combining these two results, we obtain a construction of QPKE from specific OWGAs and from TCFs.

\subsubsection{Eavesdropper-Blind Remote State Preparation} \label{intro:eb-rsp}
We introduce a new primitive called \emph{eavesdropper-blind remote state preparation}, which is a variant of RSP between a classical verifier $V$ and a quantum prover $P$. Loosely speaking, it ensures that, at the end of the protocol, $P$ holds a single-qubit state from a very large family of states that we call \emph{root of unity states}, while $V$ efficiently learns the classical description of that state. The security guarantee ensures that any QPT eavesdropper given the transcript of the classical communication between an honest $V$ and an honest $P$ learns no information about the prepared quantum state.\footnote{We emphasize that the definition of eavesdropper blindness differs from the well-known notion of \emph{honest-but-curious blindness}: in our setting, we require blindness only against a malicious third party that has access to the transcript generated by the honest verifier and honest prover, rather than against the honest prover participating in the RSP protocol itself. The key distinction is therefore that the eavesdropper does not obtain access to the internal state of the honest prover, and hence also not to the prepared state, making this definition strictly weaker.}

\begin{pdefinition}[EB-RSP, Informal]
    An \emph{eavesdropper-blind remote state preparation} protocol is an interactive protocol between a PPT verifier $V$ and a QPT prover $P$ that satisfies the two properties:
     \begin{itemize}
         \item \emph{(Correctness)} At the end of the protocol, the honest prover $P$ holds the state
         \[
         \ket{+_\theta}
         := \dfrac{1}{\sqrt{2}} (\ket{0} + e^{i \theta} \ket{1}),
         \]
         for some angle $\theta \in \Theta_N := \braces{k \cdot \tfrac{2\pi}{N} \mid 0 \leq k \leq N-1}$. On the other hand, the verifier $V$ holds the classical description of the state, namely $\theta$.
         
         \item \emph{(Eavesdropper Blindness)}
         After the execution of the EB-RSP protocol (between the honest verifier $V$ and the honest prover $P$), no QPT eavesdropper $\Ac$ 
         that receives the transcript of the execution can distinguish the real $\theta$ from a uniformly random sample drawn from $\Theta_N$.
       \end{itemize}
\end{pdefinition}

We also consider a \emph{two-message} version of this definition, in which the communication consists of a message from the verifier followed by a message from the prover.

Our main theorem is that specific OWGAs are sufficient to realize two-message EB-RSPs. Before stating this result, we first introduce the definitions of a group action and the one-wayness property.

\begin{pdefinition}[Group Action]
    Let $G$ be an arbitrary group and $X$ be a non-empty set. A \emph{group action} is a map
    \begin{align*}
        \star : G \times X &\to X \\
        (g, x) &\mapsto g \star x,
    \end{align*}
    such that the following conditions hold:
    \begin{enumerate}
        \item $1 \star x = x$ for all $x \in X$, where 1 is the identity element of $G$;

        \item $g \star (h \star x) = (gh) \star x$ for all $g, h \in G$ and all $x \in X$. 
    \end{enumerate}
\end{pdefinition}

Note that we use multiplicative notation for statements and definitions concerning arbitrary, potentially non-abelian groups. However, we consistently use additive notation for the abelian groups appearing in the construction.

We call a group action \emph{free} (or \emph{semiregular}) if the condition $g \star x = x$ for some $x \in X$ already implies that $g = 1$ (which intuitively corresponds to injectivity). Moreover, to place this mathematical object on cryptographic foundations, we assume, loosely speaking, that we work with finite objects and that all main operations on the group can be carried out efficiently. The one-wayness property states that, if a QPT adversary is given $(x, s \star x)$, where $x \leftarrow X$ and $s \leftarrow G$, it is hard to output an $s'$ such that $s' \star x = s \star x$; that is, it is hard to find a preimage of $s \star x$ under the group action with respect to $x$.

Equipped with these definitions, we can now state our main result:

\begin{ptheorem}[OWGA implies EB-RSP, Informal]
    The existence of a free OWGA $\star : G \times X \to X$, where $G = \Z_p^\lambda$ for a prime $p$, implies the existence of a two-message EB-RSP.
\end{ptheorem}

Let us first describe the protocol at a high level, and afterwards provide intuitive explanations for why correctness and security hold. At a high level, our construction proceeds as follows:

\begin{enumerate}
    \item \textbf{Verifier}: Samples $x_0 \leftarrow X$ and $r, s \leftarrow G$, and computes $x_1 := (-s) \star x_0$. Sends $(x_0, x_1, r)$ to the prover.
    
    \item \textbf{Prover}:
    \begin{enumerate}
        \item Prepares the state $\sum_{b \in \bits, g \in G} \ket{b, g}$.
        
        \item Given $x_0$ and $x_1$, computes the state:
        \[
        \sum_{b \in \bits, g \in G} \ket{b, g, g \star x_b}.
        \]
        
        \item Measures the third register in the standard basis and uses $r$ to prepare the state:
        \[
        \ket{0, g_0, \langle g_0, r \rangle} + \ket{1, g_0 + s, \langle g_0 + s, r \rangle}.
        \]

        \item Measures the last two registers in the QFT basis, resulting in outcomes $g$ and $z$. Sends $(g, z)$ to verifier.

        \item Outputs $\ket{+_\theta}$, where $\theta := \langle s, z \cdot r + g \rangle$.
    \end{enumerate}
        
    \item \textbf{Verifier}: Uses $s, r, z, g$ to compute $\theta$.
\end{enumerate}

\smallskip\noindent\textbf{Correctness.}
Let us start by explaining why this procedure intuitively yields our desired quantum state. We basically rely on two properties of the free OWGA: the freeness and the algebraic properties of the group action. After measuring the last register in step (b) of the prover, the state collapses to a \emph{claw state} of the form
\[
\ket{0, g_0} + \ket{1, g_0 + s},
\]
because the group action is, by assumption, free (which translates roughly to being injective in some sense). Later, in step (d), we measure this quantum state in the Fourier basis to push the appearing terms in the state into a relative phase. Because of the algebraic properties of the group action, unknown parameters like $g_0$ cancel out, and we are left with $\theta := \langle s, z \cdot r + g \rangle$, which can be efficiently computed by the verifier.

\bigskip\noindent\textbf{Security.}
Let us now move on to the intuitive explanation of the security. This follows relatively immediately, since $g$ is a uniform element in $G$ and independent of everything (as the prover behaves honestly). Thus, $\theta$ essentially consists of a secret $s$, unknown to the adversary, and a uniformly random quantity $z \cdot r + g$, which is independent of everything. This is exactly the setting of the quantum version of the Goldreich-Levin theorem for large fields given in \cref{thm:gl-quantum}, which states that this quantity is computationally indistinguishable from uniform.

\bigskip\noindent\textbf{Protocol Simplification.}
After completing the first version of the current paper, we realized that the presented protocol can be further simplified, as also pointed out independently by an anonymous Asiacrypt reviewer. Concretely, the protocol can be simplified by removing $r$ and the measurement outcome $z$ from the construction altogether, which has the same effect as setting $r = 0$, i.e., the all-zero string. 
Completeness remains unchanged, as it holds for every value of $r$.
For eavesdropper blindness, observe that, for any fixed verifier randomness $r$ and any outcome of the inner product register measurement, the Fourier outcome $g$ is uniformly distributed over $G$. In particular, $g$ is uniform and independent of the verifier's first message. And now one can use the Goldreich-Levin argument for $\langle s, g\rangle$ as before.

\bigskip\noindent\textbf{Family of Output States.}
Our EB-RSP is more general than most existing protocols in terms of the family of states it can remotely prepare. We consider $N$ states equally spaced on the $(X,Y)$-plane of the Bloch sphere, which we call \emph{root of unity states}, since they are $\ket{+}$ states with relative phases given by the $N$-th roots of unity. Most prior RSP protocols cover only special cases of this family, such as BB84 states \cite{C:BarKhu25} and generalized BB84 states \cite{AC:CCKW19,FOCS:GheVid19,EPRINT:AMMW22,C:BKMSW25,ITCS:Zhang25}.

Moreover, the presented protocol is for states with $N = p$ being a prime, but it also extends to $N = p_1 \cdots p_m$ for pairwise distinct primes $p_i$ by combining the above EB-RSP in parallel. View the presented EB-RSP as a subroutine in which the verifier inputs $p$ and the prover inputs a one-qubit state. At the end of the subroutine, a relative phase is included on the prover's input qubit, while the verifier knows this relative phase. Then run the subroutine for all $p_i$, yielding a state $\ket{+_{\theta}}$ with $\theta = \theta_1 + \ldots + \theta_m$ and $\theta_i \in \Theta_{p_i}$. Note that, in the end, $\theta = \theta_1 + \ldots + \theta_m \in \Theta_N$ by simply extending the fractions, which gives the desired EB-RSP for more general $N$. Security follows from the independence of the subroutines, the security for each $\theta_i$, and the distinctness of the primes. However, how to extend this idea to powers of primes remains unclear.

\bigskip\noindent\textbf{No requirement of trapdoor.}
As we can see, the EB-RSP protocol presented here does not require the verifier $V$ to use a trapdoor to compute the classical description of the quantum state, i.e. $V$ does not have to invert a function using secret information, since we exploit only the algebraic properties of the group action to cancel out parameters unknown to the verifier. This ensures that only parameters known to the verifier appear in the quantum state while still maintaining security. To the best of our knowledge, this is the first RSP-type primitive with useful applications that does not rely on any trapdoor. The techniques used here could enable the blind preparation of more specific quantum states with stronger security properties without relying on a trapdoor primitive, thereby taking a step toward reducing the cryptographic assumptions underlying many protocols, as RSP protocols form the backbone of many quantum cryptographic schemes. This is an important goal, as reducing complexity assumptions remains a major ongoing work in cryptography.

\bigskip
This new primitive will ultimately be used in \cref{sec:qpke} to construct a quantum public-key encryption (QPKE) scheme with classical public keys in a black-box fashion. We also want to highlight that since our EB-RSP includes a wide range of output states, together with a relatively weak security definition, we can hope for relatively simple constructions from different cryptographic assumptions, allowing the QPKE to be instantiated from a wider range of assumptions. It might also be the case that many known RSPs already fall into our category of EB-RSP, such as the protocol presented in \cite{C:BKMSW25} (described in \cref{sec:appendix}), after a few suitable modifications, yielding the following result:

\begin{ptheorem}[TCF implies EB-RSP, Informal]
    The existence of a plain TCF implies the existence of a two-message EB-RSP.
\end{ptheorem}

\subsubsection{Quantum Public-Key Encryption with Classical Public Keys}
Our main result about QPKE is the black-box construction from two-message EB-RSP.

\begin{ptheorem}[EB-RSP implies QPKE, Informal]
    The existence of a two-message EB-RSP implies the existence of a QPKE scheme with classical public keys and quantum ciphertexts.
\end{ptheorem}

Before presenting the high-level description of the construction, we list the corollaries obtained from the above theorem by combining it with the previous theorems.

\begin{pcorollary}[OWGA implies QPKE, Informal]\label{cor:owga-qpke-informal}
    The existence of a free OWGA $\star : G \times X \to X$, where $G = \Z_p^\lambda$ for a prime $p$, implies the existence of a QPKE scheme with classical public keys and quantum ciphertexts.
\end{pcorollary}

\begin{pcorollary}[TCF implies QPKE, Informal]\label{cor:tcf-qpke-informal}
    The existence of a plain TCF implies the existence of a QPKE scheme with classical public keys and quantum ciphertexts.
\end{pcorollary}

\begin{premark}\label{remark:relation-HMY}
The assumption-level conclusions of Corollaries~\ref{cor:owga-qpke-informal} and~\ref{cor:tcf-qpke-informal} can also be obtained by specializing the framework of \cite{EC:HhaMorYam23}.
Concretely, for the free group actions considered here, define
$f_b(h):=h\star x_b \text{, for }b\in\{0,1\}$.
Freeness implies that each $f_b$ is injective, while one-wayness of the group action implies that the pair $(f_0,f_1)$ is claw-free. Since injective functions are collapsing, \cite{EC:HhaMorYam23} shows that claw-freeness together with collapsingness implies conversion hardness, which in turn suffices for constructing an IND-CPA-secure QPKE scheme with classical public keys and quantum ciphertexts.
Likewise, a TCF gives a swap-trapdoor function pair: trapdoor inversion recovers the unique matching preimage in the other branch, and the branch functions are injective and claw-free.
Consequently, the two assumption-level QPKE corollaries above should be viewed as alternative modular derivations rather than new feasibility results relative to \cite{EC:HhaMorYam23}.
\end{premark}

Let us now describe how the QPKE scheme from two-message EB-RSP looks at a high level, and afterwards give intuitive explanations of why correctness and security hold. For this, we first present the main ingredients of the definition of a QPKE scheme with classical public keys for classical one-bit messages. How to extend the message space is explained in \cref{subsec:qpke-definition}. Such a QPKE scheme consists of the algorithms $(\keygen, \enc, \dec)$:
\begin{itemize}
    \item $\keygen(1^\lambda) \rightarrow (\pk, \sk)$. This is a PPT algorithm that takes the security parameter $1^\lambda$ as input and outputs a classical public key $\pk$ and a classical secret key $\sk$.

    \item $\enc(\pk, b) \rightarrow ct$. This is a QPT algorithm that takes a public key $\pk$ and a message $b \in \bits$ as input and outputs a quantum ciphertext $ct$.

    \item $\dec(\sk, ct) \rightarrow b'/\perp$. This is a QPT algorithm that takes a secret key $\sk$ and a ciphertext $ct$ as input and outputs a message $b' \in \bits$ or $\perp$.
\end{itemize}

For security, we follow the standard IND-CPA security definition. Note that our main theorem concerns the construction of QPKE from \emph{any} two-message EB-RSP. For the sake of understanding, we focus in this section on the concrete protocol presented above and give a simplified version here.

At a high level, our construction works as follows.
\begin{itemize}
    \item $\keygen$: We define the public key $\pk := (x_0, x_1, r)$ to be the first message sent by the verifier, and the secret key $\sk := s$ to be the internal state of the verifier after sending the first message. It is essentially the information that the verifier does not reveal and later uses to compute the secret angle.

    \item $\enc$: To encrypt a bit $b \in \bits$, one goes through all the steps the prover performs in the EB-RSP, after receiving the message from the verifier, until obtaining a quantum state $\ket{+_\theta}$ and the two outcomes $(z,g)$. One then applies a unitary $U_b := R_z(b \cdot \floor{\tfrac{N}{2}} \cdot \tfrac{2\pi}{N})$, which encodes the information about $b$, to the state, and outputs $ct := (U_b \ket{+_\theta}, z, g)$ as the ciphertext. Thus, the ciphertext consists of the messages sent by the prover in the EB-RSP together with a unitary, depending on $b$, applied to the output state.
    
    \item $\dec$: One uses the secret key $\sk = s$ and the classical information within the ciphertext, to recover $\theta$. One then applies the unitary $R_z(-\theta)$ to the quantum state, yielding the state $U_b \ket{+}$. Lastly, one measures this state in the Hadamard basis to obtain the initial message $b$.
\end{itemize}

\smallskip\noindent\textbf{Correctness.} As mentioned, we consider only a simplified version here in order to understand the mechanics of the QPKE scheme. In the decryption phase, we measure the state $U_b \ket{+}$ in the Hadamard basis. This state is precisely $\ket{+}$ for $b = 0$, and is a state very close to $\ket{-}$ for $b = 1$. Hence, correct decryption occurs with relatively high probability. In the actual scheme, this probability is amplified by encrypting many states and taking the majority vote of the outcomes.

\bigskip\noindent\textbf{Security.} For security, the adversary essentially has access to the public key $\pk$ and the ciphertext $ct = (U_b \ket{+_\theta}, z, g)$. We may even imagine giving the adversary the classical description of the state, which can only increase the adversary's advantage, yet it remains negligible, as we will see. This classical description can essentially be written as $\theta + b \cdot \floor{\tfrac{N}{2}} \cdot \tfrac{2\pi}{N}$. Since $\theta$ is computationally indistinguishable from uniform due to the eavesdropper blindness of the EB-RSP, this classical information hides $b$, giving the adversary no non-negligible advantage in guessing $b$.

\bigskip\noindent\textbf{On the Security of the EB-RSP for QPKE.} Our EB-RSP security definition considers the full angle $\theta = k \cdot \tfrac{2\pi}{N}$ with $k \in \Z_N$ to be computationally indistinguishable from uniform. When $N$ is a power of $2$, this means that every bit of $k$ is hidden.

But note that for IND-CPA security, we only need the most significant bit of $k$ to be computationally indistinguishable, since our encryption essentially flips the most significant bit of $k$ within the quantum state when $N$ is a power of $2$. For general $N$, this translates to effectively hiding the single bit of information corresponding to whether $0 \leq k < \floor{\tfrac{N}{2}}$ or $\floor{\tfrac{N}{2}} \leq k < N$, that is, whether $\theta$ lies on the upper half or the lower half of the unit circle. Thus, requiring only that this information be hidden would be sufficient to ensure IND-CPA security. This is exactly equivalent to hiding the most significant bit of $k$ when $N$ is a power of $2$. An interesting fact is that all other known RSPs do it the other way around: they try to hide all but the most significant bit (since it is impossible to hide everything while allowing the prover itself to be malicious).

\subsection{Discussions}\label{subsec:discussions}
In this section, we discuss related work and outline several future directions.

\bigskip\noindent\textbf{Related Work.} A long line of work on remote state preparation has focused on achieving strong notions of blindness and verifiability from trapdoor-based cryptographic assumptions, most notably variants of trapdoor claw-free functions \cite{AC:CCKW19,FOCS:GheVid19,EPRINT:AMMW22,C:GupVai24,C:BarKhu25,C:BKMSW25,ITCS:Zhang25}. In contrast, our work studies whether weaker forms of remote state preparation may still suffice for meaningful cryptographic applications while admitting constructions from qualitatively different assumptions. More concretely, our work introduces EB-RSP, a weaker RSP-type primitive satisfying only blindness against eavesdroppers. Conceptually, our approach differs from prior RSP constructions in two ways. First, we relax the blindness requirement compared to standard malicious-server blindness notions. Second, rather than relying on trapdoor cryptographic primitives, our construction exploits algebraic properties of one-way group actions to obtain a classical description of the prepared state.

Group actions and related algebraic assumptions have previously appeared in several quantum cryptographic constructions. In particular, \cite{EC:HhaMorYam23} constructs QPKE with classical public keys and quantum ciphertexts through the abstraction of conversion-hard swap-trapdoor function pairs. Their general group-action instantiation is analyzed under a \emph{pseudorandomness} assumption to obtain the QPKE construction. 
However, as discussed in \cref{remark:relation-HMY}, it can also be shown that their framework also yields the OWGA-to-QPKE consequence for the free group actions considered here: freeness makes the associated branch functions injective and hence collapsing, while one-wayness of the action implies claw-freeness; their conversion-hardness result then yields QPKE.
Our work pursues a complementary objective. We introduce two-message EB-RSP as an abstract state-preparation functionality that already suffices for QPKE, establish a generic black-box transformation from any two-message EB-RSP to QPKE and construct such an EB-RSP from free OWGAs without relying on a trapdoor. This modular route separates the construction of the remote state preparation primitive from its use in encryption and, to the best of our knowledge, yields the first trapdoor-free RSP-type primitive with a concrete cryptographic application.

The idea of constructing QPKE from RSP has also appeared in prior work. In particular, \cite{C:BarKhu25} showed how to construct QPKE with classical public keys and classical ciphertexts from RSPs based on dual-mode trapdoor claw-free functions. In contrast, our work shows that even weaker forms of RSP already suffice for constructing QPKE with classical public keys, albeit with quantum ciphertexts. Furthermore, we observe that existing RSP constructions appear naturally adaptable to our two-message EB-RSP notion. We demonstrate this explicitly for a concrete TCF-based RSP construction, immediately yielding a construction of QPKE from plain TCFs.

\bigskip\noindent\textbf{Future Directions.} Our work suggests several directions for future research. A first conceptual direction is to better understand the landscape of weaker RSP-type primitives and their cryptographic applications. In this work, we showed that even the relatively weak notion of EB-RSP already suffices to construct QPKE with classical public keys and quantum ciphertexts. This raises the broader question of which security properties of remote state preparation are truly necessary for various quantum cryptographic tasks. For instance, it would be relevant to identify additional applications that can already be realized from the security notion of eavesdropper-blindness, particularly in the QCCC model. 

Another interesting direction is to investigate whether similar ideas can be extended to obtain stronger notions of remote state preparation, such as blind or verifiable RSP, from trapdoor-free assumptions.

\noindent\emph{Towards stronger RSP and proofs of quantumness.} A natural direction is to investigate whether the algebraic phase-generation mechanism developed here can be combined with additional techniques to obtain stronger notions of remote state preparation, such as blindness against a malicious prover or verifiable RSP, while retaining trapdoor-free assumptions. Furthermore, several RSP and proof-of-quantumness protocols exploit closely related tools, including coherent claw states and measurements in complementary or Fourier bases. Therefore, a closely related target would be the construction of proofs of quantumness from group-action-based assumptions.

Finally, several technical questions remain open. One natural direction is whether our OWGA-based construction can be generalized beyond the current classes of groups considered in this work. Another is whether similar techniques could eventually yield QPKE schemes with classical public keys and classical ciphertexts, as in \cite{C:BarKhu25}, while still avoiding trapdoor assumptions.

\bigskip\noindent\textbf{Disclosure on the use of AI.}
The authors declare that no AI tools were used at any stage of the research process. All intellectual and written contributions to the study were made by the authors without AI assistance.

%% file: sec-prelim.tex
\section{Preliminaries}\label{sec:prelim}
We denote the security parameter by $\lambda \in \N$. A function $\negl$ is called \emph{negligible} if it vanishes faster than the absolute value of any inverse polynomial. We denote by $\omega_n = e^{2\pi i/n}$ the $n$-th root of unity. We define the set
\[
\Theta_N 
:= \braces{k \cdot \dfrac{2\pi}{N} \mid 0 \leq k \leq N-1}
\]
for any positive integer $N$, as we will work with it extensively. We also define the rotated $\ket{+}$ state
\[
\ket{+_\theta}
:= \dfrac{1}{\sqrt{2}} (\ket{0} + e^{i \theta} \ket{1}),
\]
which is defined for any $\theta \in \R$. Lastly, for any $\theta \in \R$, we define the \emph{rotation operator about the $z$-axis} with respect to the Bloch sphere as follows:
\begin{align*}
	R_z(\theta) 
	&= \begin{pmatrix}
			e^{-i \theta/2} & 0 \\
			0 & e^{i \theta/2}
	\end{pmatrix},
\end{align*}
which is equivalent to $\left(\begin{smallmatrix}
1 & 0 \\
0 & e^{i\theta}
\end{smallmatrix}\right)$ up to a global phase.

\subsection{One-Way Group Actions}\label{subsec:owga}
We briefly lay the mathematical foundation of this work by recalling the definition of an abstract group action in the mathematical sense. We then place this mathematical object on cryptographic grounds by requiring that certain operations are efficient, together with equipping it with a cryptographic assumption: the \emph{one-wayness}, as in the seminal work of Brassard and Yung \cite{C:BraYun90}.

Let us now fix the notation used for groups throughout this work. For abelian groups $G$, we use additive notation to denote the group operation, i.e., $g + h$ for elements $g,h \in G$. Moreover, the neutral element is denoted by $0$, and the inverse of an element $g \in G$ by $-g$. For general groups $G$ that may not be abelian, we use multiplicative notation to denote the group operation, i.e., $g \cdot h$ or $gh$ for elements $g,h \in G$. Moreover, the neutral element is denoted by $1$, and the inverse of an element $g$ by $g^{-1}$. In the remaining of the paper we will use multiplicative notation for statements and definitions concerning arbitrary, potentially non-abelian groups. However, we consistently use additive notation for the abelian groups appearing in the constructions later on.

\begin{pdefinition}[Group Action]
    Let $G$ be an arbitrary group and $X$ be a non-empty set. A \emph{group action} is a map
    \begin{align*}
        \star : G \times X &\to X \\
        (g, x) &\mapsto g \star x,
    \end{align*}
    such that the following conditions hold:
    \begin{enumerate}
        \item $1 \star x = x$ for all $x \in X$;

        \item $g \star (h \star x) = (gh) \star x$ for all $g, h \in G$ and all $x \in X$. 
    \end{enumerate}
\end{pdefinition}

We now introduce some common terminology used when working with group actions. Let $x \in X$ be any element. We call 
\[
G_x := \braces{g \in G \mid g \star x = x}
\]
the \emph{stabilizer group} of $x$. The group action is called \emph{free} (or \emph{semiregular}) if the statement that $g \star x = x$ for some $x \in X$ already implies that $g = 1$. Or in other words, the stabilizer group of each $x$ is trivial. 

Next, we place this mathematical object on cryptographic grounds. To do this, we essentially assume that we work with finite objects and that all important operations we need to perform can be carried out efficiently. We follow (more or less) the definition given in \cite[Definition 4.13]{EC:HhaMorYam23}, but we do not require all the properties they list.

\begin{pdefinition}[Effective Group Action]
    A group action $\star : G \times X \to X$ is called an \emph{effective group action (EGA)} if the following properties hold:
    \begin{enumerate}[label=\textnormal{(\roman*)}]
        \item $G$ and $X$ are finite.

        \item There are classical deterministic polynomial-time algorithms that compute the multiplication $g \cdot h$, the inverse $g^{-1}$, and the group action $g \star x$ for all $g,h \in G$ and $x \in X$.

        \item There are PPT algorithms that sample uniformly at random from $G$ and $X$.

        \item There is a QPT algorithm that generates a uniform superposition over all elements $g \in G$.
\end{enumerate}
\end{pdefinition}

Such EGAs can now be endowed with cryptographic hardness assumptions, and researchers always come up with new hardness assumptions for cryptographic group actions to prove the security of their protocols. We will focus on one of the weakest such assumptions among all assumptions currently in circulation.

\begin{pdefinition}[One-Way Group Action]
    An effective group action $\star : G \times X \to X$ is called an \emph{one-way group action (OWGA)} if, for any QPT adversary $\Ac$, we have
    \begin{align*}
    \Pr[s \leftarrow G \\ x \leftarrow X]{s' \star x = s \star x \, : \, s' \leftarrow \Ac(1^\lambda, x, s \star x)}
    = \negl(\lambda).
    \end{align*}
\end{pdefinition}

Note that for a free group action, the QPT adversary has to find $s$ itself.

\subsection{Quantum Fourier Transform over Finite Abelian Groups}\label{subsec:qft}
Let $G$ be a finite abelian group. By the well-known Fundamental Theorem of Finite Abelian Groups, we know that every finite abelian group $G$ is isomorphic to a direct sum of cyclic groups
\[
\Z_{N_1} \times \ldots \times \Z_{N_k},
\]
for some non-negative integer $k$ and specific $N_i$, each of which is a power of a (not necessarily distinct) prime. For simplicity, we assume that $G$ equals such a group, that is, $G = \Z_{N_1} \times \ldots \times \Z_{N_k}$. For an element $g \in G$, we denote its $i$-th component by $g_i$, that is, $g = (g_1, \ldots, g_k)$. Moreover, we define the map $\chi : G \times G \to \C$ by
\[
\chi(h, g)
:= \omega_{N_1}^{h_1 g_1} \cdots \omega_{N_k}^{h_k g_k}.
\]
We are now ready to define the quantum Fourier transform.

\begin{pdefinition}[Quantum Fourier Transform over Finite Abelian Groups]
    The \emph{quantum Fourier transform (QFT) over finite abelian groups} $G$ is the unitary operator $\QFT_G$ defined by
    \[
    \QFT_G \ket{g}
    = \dfrac{1}{\sqrt{\abs{G}}} \sum_{h \in G} \chi(h, g) \ket{h}.
    \]
\end{pdefinition}
Note that the QFT decomposes as
\[
\QFT_G 
= \QFT_{\Z_{N_1}} \otimes \ldots \otimes \QFT_{\Z_{N_k}},
\]
and can therefore be implemented efficiently.

Next, we introduce some notation to make the expressions appearing later in this work more concise. Define $M := \lcm(N_1, \ldots, N_k)$ to be their least common multiple. For each $1 \leq i \leq k$, set $R_i := M / N_i$. Consider the following two maps:
\begin{align*}
    f_1 : G \times G &\to G \\
    (h, g) &\mapsto (h_1 g_1, \ldots, h_k g_k),
\end{align*}
where each multiplication is performed with respect to the corresponding modulus, and
\begin{align*}
    f_2 : G &\to \Z_M \\
    e &\mapsto \sum_{i = 1}^{k} e_i R_i,
\end{align*}
where the multiplication and summation are performed modulo $M$. Both maps are clearly well defined. Lastly, we define the map
\[
\langle \cdot, \cdot \rangle : G \times G \to \Z_M
\]
via $\langle \cdot, \cdot \rangle = f_2 \circ f_1$, which is the composition of the two maps defined above. It is easy to see that $\langle \cdot, \cdot \rangle$ is symmetric and homomorphic in both components; that is, $\langle h, g \rangle = \langle g, h \rangle$ and $\langle h + g, e \rangle = \langle h, e \rangle + \langle g, e \rangle$. Now, to draw the connection to $\chi$, we observe that
\[
\chi(h, g) 
= \omega_{N_1}^{h_1 g_1} \cdots \omega_{N_k}^{h_k g_k} 
= \omega_{M}^{h_1 g_1 R_1} \cdots \omega_{M}^{h_k g_k R_k} 
= \omega_{M}^{\langle h, g \rangle}.
\]
Hence, the newly established notation allows us to express $\chi$ more concisely. Moreover, this immediately shows that $\chi$ is also symmetric in $h$ and $g$, that is, $\chi(h, g) = \chi(g, h)$, and bilinear; that is, $\chi(h + g, e) = \chi(h, e) \cdot \chi(g, e)$, with a similar property holding for the second component. In this work, we focus on the case where $N_1 = \ldots = N_k = p$ for a prime $p$. Consequently, $M = p$, and every operation is performed modulo $p$.

\subsection{Quantum Goldreich-Levin Theorem for Large Fields}\label{subsec:gl-theorem}
We recall generalized Goldreich-Levin theorems over large fields, one for classical adversaries and one for quantum adversaries. We begin with the classical version.

\begin{ptheorem}[{Classical Goldreich-Levin Theorem for Large Fields \cite[Theorem 1]{TCC:DGKPV10}}]\label[ptheorem]{thm:gl-classical}
    Let $p$ be prime, and let $H$ be an arbitrary subset of $\Z_p$. Let $f : H^\lambda \to \bits^*$ be any (possibly randomized) function. If there is a distinguisher $\Dc$ that runs in time $t$ such that
    \begin{align*}
        \abs{
        \Pr[x \leftarrow H^\lambda \\ y \leftarrow f(x) \\ r \leftarrow \Z_p^\lambda]{\Dc(y, r, \langle r, x \rangle) = 1}
        - \Pr[x \leftarrow H^\lambda \\ y \leftarrow f(x) \\ r \leftarrow \Z_p^\lambda \\ u \leftarrow \Z_p]{\Dc(y, r, u) = 1}
        }
        = \varepsilon,
    \end{align*}
    then there is an inverter $\Ac$ that runs in time $t' = t \cdot \poly(\lambda, \abs{H}, 1/\varepsilon)$ such that 
    \[
    \Pr[x \leftarrow H^\lambda \\ y \leftarrow f(x)]{\Ac(y) = x}
    \geq \dfrac{\varepsilon^3}{512 \cdot \lambda \cdot p^2}. 
    \]
\end{ptheorem}

Now, we want to obtain a quantum version of the above theorem. To do this, we use the same idea explained in \cite{TCC:AnaPorVai23} by applying \cite[Theorem 4.1]{TCC:AnaPorVai23}, an adapted version of \cite[Theorem 7.1]{C:BitBraKal22}, to \cref{thm:gl-classical}. This converts classical reductions into post-quantum reductions.

\begin{ptheorem}[Quantum Goldreich-Levin Theorem for Large Fields]\label[ptheorem]{thm:gl-quantum}
    Let $p$ be prime, and let $H$ be an arbitrary subset of $\Z_p$. Let $\Phi: \Lc(\Hc_p^{\otimes \lambda}) \rightarrow \Lc(\Hc_{\mathsf{Aux}})$ be any CPTP map with auxiliary system $\Hc_{\mathsf{Aux}}$, where $\Hc_p$ is $p$-dimensional. If there is a quantum distinguisher $\Dc$ that runs in time $t$ such that
    \begin{align*}
        \abs{
        \Pr[x \leftarrow H^\lambda \\ \aux \leftarrow \Phi({\ket{x}\hspace{-0.4mm}\bra{x}}) \\ r \leftarrow \Z_p^\lambda]{\Dc(\aux, r, \langle r, x \rangle) = 1}
        - \Pr[x \leftarrow H^\lambda \\ \aux \leftarrow \Phi({\ket{x}\hspace{-0.4mm}\bra{x}}) \\ r \leftarrow \Z_p^\lambda \\ u \leftarrow \Z_p]{\Dc(\aux, r, u) = 1}
        }
        = \varepsilon,
    \end{align*}
    then there is a quantum extractor $\Ec$ that runs in time $t' = t \cdot \poly(\lambda, \abs{H}, 1/\varepsilon)$ such that 
    \[
    \Pr[x \leftarrow H^\lambda \\ \aux \leftarrow \Phi({\ket{x}\hspace{-0.4mm}\bra{x}})]{\Ec(\aux) = x}
    \geq \poly(\varepsilon, 1/\lambda, 1/p).
    \]
\end{ptheorem}

We will work with this theorem in the setting where $p$ is polynomial in $\lambda$, $H = \Z_p$, $\Phi$ computes the output of a quantumly secure one-way function $f$, and $\Dc$ is a QPT algorithm. In this setting, the advantage $\varepsilon$ of $\Dc$ is negligible, i.e., the hardcore predicate $\langle r, x \rangle$ looks uniformly random from the adversary's view.

%% file: sec-rsp.tex
\section{Eavesdropper-Blind Remote State Preparation}\label{sec:rsp}
In this section, we present a new type of \emph{remote state preparation} (RSP) protocol for \emph{root of unity states}, which generalize the BB$84$ states,
\[
\ket{+_\theta}
:= \dfrac{1}{\sqrt{2}} (\ket{0} + e^{i \theta} \ket{1}),
\]
where
\[
\theta \in \Theta_N 
:= \braces{k \cdot \dfrac{2\pi}{N} \mid 0 \leq k \leq N-1},
\]
from free OWGAs associated with certain groups. We consider a weaker security notion for RSP than the standard ones, which we call \emph{eavesdropper blindness}, shifting the focus from the server to an external eavesdropper. We emphasize that the definition of eavesdropper blindness differs from the well-known notion of \emph{honest-but-curious blindness}: in our setting, we require blindness only against a malicious third party that has access to the transcript generated by the honest verifier and honest prover, rather than against the honest prover participating in the RSP protocol itself. The key distinction is therefore that the eavesdropper does not obtain access to the internal state of the honest prover, and hence also not to the prepared state, making this definition strictly weaker.

A \emph{two-message version} of this new primitive will ultimately be used in \cref{subsec:qpke-construction} to construct a quantum public-key encryption (QPKE) scheme with classical public keys in a black-box fashion. Although this section provides a concrete instantiation based on certain OWGAs, we later adopt a modular approach to the QPKE construction that relies only on the abstract definition of a two-message eavesdropper-blind RSP (EB-RSP) protocol, rather than on any specific instantiation. This allows any two-message EB-RSP protocol to be used within the QPKE scheme. Nevertheless, the protocol presented here may be of independent interest. In particular, we construct this two-message EB-RSP from a cryptographic primitive in such a way that the verifier does not require a trapdoor, since the construction exploits only the algebraic properties of the group action. To the best of our knowledge, this is the first RSP-type primitive that does not rely on any trapdoor. The techniques used here could enable the blind preparation of more specific quantum states with stronger security properties without relying on a trapdoor primitive, thereby taking a step toward reducing the cryptographic assumptions underlying many protocols, as RSP protocols form the backbone of many quantum cryptographic schemes. This is an important goal, as reducing complexity assumptions remains a major ongoing work in cryptography.

\subsection{Definition}\label{subsec:rsp-definition}
We start by providing a formal definition of this new primitive and will later explain in more detail its functionality and how it compares to other common RSP definitions.

\begin{pdefinition}[Eavesdropper-Blind Remote State Preparation]\label[pdefinition]{def:eb-rsp}
    An \emph{eavesdropper-blind remote state preparation (EB-RSP)} protocol consists of a pair of interactive algorithms $(V, P)$, with the security parameter in unary $1^\lambda$ as input: A classical probabilistic polynomial-time algorithm $V$, called the \emph{verifier}, and a quantum polynomial-time algorithm $P$, called the \emph{prover}. We require the protocol to satisfy the following properties:

    \begin{itemize}
        \item \emph{(Correctness)} The protocol terminates successfully with probability at least $1 - \emph{\textsf{negl}}(\lambda)$. Furthermore, upon successful completion, the honest prover $P$ holds the state
        \[
        \ket{+_{\theta}}
        \]
        for an angle $\theta \in \Theta_N := \braces{k \cdot \tfrac{2\pi}{N} \mid 0 \leq k \leq N-1}$, where $N = N(\lambda)$ is a positive integer that depends on $\lambda$. The verifier, on the other hand, holds the classical description of the state, i.e. $\theta$. Denote the transcript of the classical communication by $\trans$.
        
        \item \emph{(Eavesdropper Blindness)} Consider the following experiment $\mathrm{Exp}(1^\lambda, V, P, \Ac)$ played between an honest verifier $V$, an honest prover $P$, and an eavesdropper $\Ac$.\footnote{Note that, for our security definition, in the experiment we consider indistinguishability-type security for $\theta$. One might think that a search-type security definition, in which the adversary is asked to output the correct $\theta$ from the search space $\Theta_N$, would generally be sufficient, but this is not the case. In fact, if the search space, parameterized here by $N$, is polynomial in the security parameter $\lambda$, then the two definitions are equivalent. However, for general $N$, and in particular when $N$ is exponential in $\lambda$, indistinguishability-type security is strictly stronger and necessary for our QPKE construction. Loosely speaking, search-type security may reveal bits of $\theta$ that must remain hidden from the eavesdropper in the QPKE construction.}
        \begin{itemize}
            \item $V$ and $P$ engage in the interactive EB-RSP protocol. If the protocol does not terminate successfully, the experiment outputs $1$ with probability $\tfrac{1}{2}$ and $0$ with probability $\tfrac{1}{2}$. If the protocol does terminate successfully, it continues as follows:

            \item Let $\ket{+_{\theta}}$ be the output state held by $P$, and $\theta$ the output of $V$. Send the EB-RSP transcript $\trans$ to $\Ac$.
            
            \item Flip a coin $c \leftarrow \{0,1\}$. If $c=0$, set $\theta_{\chal} := \theta$, otherwise sample $\theta_{\chal} \leftarrow \Theta_N$ uniformly at random. Send $\theta_{\chal}$ to $\Ac$, who returns a bit $c'$.
            
            \item The experiment outputs $1$ if $c' = c$ and $0$ otherwise.
        \end{itemize}
        We require that for all QPT adversaries $\Ac$ there exists a negligible function $\negl$ such that for all sufficiently large $\lambda \in \mathbb{N}$:
        \[
        \abs{\Pr{\mathrm{Exp}(1^\lambda, V, P, \Ac)=1} - \frac{1}{2}}
        \leq \negl(\lambda).
        \]
    \end{itemize}
    We call it a \emph{two-message EB-RSP} if it additionally consists of exactly two messages: one from the verifier, followed by one from the prover. In this case, we use the following notation to describe the algorithms within the protocol:
    \begin{itemize}
        \item $\ebrsp.\setup(1^\lambda) \rightarrow (\msg_V, \st_V)$. The verifier takes as input the security parameter $1^\lambda$ and outputs a message $\msg_V$ and a state $\st_V$.

        \item $\ebrsp.\qcomp(\msg_V) \rightarrow (\ket{+_\theta}, \msg_P)$. The prover takes as input the verifier's message and outputs the state $\ket{+_\theta}$ and a message $\msg_P$.

        \item $\ebrsp.\dec(\st_V, \msg_P) \rightarrow \theta$. The verifier takes as input its state $\st_V$ and the prover's message $\msg_P$, and outputs $\theta$. 
    \end{itemize}
\end{pdefinition}

\smallskip\noindent\textbf{Family of Output States.}
First, our EB-RSP is more general than other known RSP protocols in terms of the family of states being remotely prepared. In our case, we consider $N$ states that lie equidistantly on the ($X$, $Y$)-plane of the Bloch sphere, which we call \emph{root of unity states}, as they correspond to the $\ket{+}$ state with a relative phase described by the $N$-th roots of unity. Most existing RSP constructions work with a special case of this family of states, including BB84 states \cite{C:BarKhu25} and generalized BB84 states \cite{AC:CCKW19,FOCS:GheVid19,EPRINT:AMMW22,C:BKMSW25,ITCS:Zhang25}. However, there is also at least one other known RSP protocol that does not fall into this category and instead prepares EPR pairs \cite{C:GupVai24}.

\bigskip\noindent\textbf{Security.}
The main difference between our EB-RSP and others lies in the security notion. Standard RSP security definitions aim to hide the angle $\theta$ from the quantum prover. Moreover, these definitions typically allow the prover to deviate arbitrarily from the protocol while still requiring that $\theta$ remains hidden, with the disadvantage of introducing an undesired \emph{byproduct bit}\footnote{The byproduct bit intuitively refers to a single bit from the entire description of $\theta$, as we will see later in the discussion about the RSP from \cite{C:BKMSW25}.}, over which the verifier has no control and for which one usually has no security guarantees.
By contrast, our security notion concerns hiding the angle only from an external third party that passively intercepts the communication between an honest verifier and an honest prover, and no undesired byproduct bit appears in this construction.
More precisely, eavesdropper blindness means that the protocol is blind only with respect to a passive third-party observer of the communication channel, and not with respect to the server itself. Crucially, this security notion is still sufficient for our QPKE application, which motivates our introduction of eavesdropper blindness.

\bigskip
Before presenting our concrete instantiation based on certain OWGAs, we emphasize that suitable instantiations already exist. For example, the RSP protocol of \cite{C:BKMSW25}, based on plain TCFs, already provides such an instantiation. A full description of the RSP is given in \cref{sec:appendix}. With only minor modifications, that protocol can be transformed into a two-message EB-RSP protocol by merging the verifier's messages into a single message and likewise merging the prover's messages. This transformation has no drawback on security. Note that their RSP prepares quantum states of the form
\[
Z^b \ket{+_\theta} 
= \ket{+_{(\theta + b\pi)}},
\]
where $\theta \in \Theta_N$ with $N = 8$, and $b$ is a single bit representing the undesired byproduct bit, while security holds only with respect to $\theta$ and not the entire angle $\theta + b\pi$, when the quantum prover is considered to be the adversary. With respect to eavesdropper blindness, however, one can verify, after a few algebraic manipulations, that from the eavesdropper's perspective, the angle $\theta + b\pi$ remains computationally indistinguishable from uniform, by an argument similar to the Goldreich-Levin argument used in their work. The intuitive reason for this is that, in the honest execution of the protocol, $b$ is independent of $\theta$, which may not be the case when the quantum prover behaves maliciously. Hence, we obtain the following result.

\begin{ptheorem}\label[ptheorem]{thm:tcf-rsp}
    The existence of a plain TCF implies the existence of a two-message EB-RSP.
\end{ptheorem}

This indicates that there may exist even more RSP protocols that can likewise be transformed into two-message EB-RSP protocols and subsequently used to construct QPKE schemes.

\subsection{Construction from One-Way Group Actions}\label{subsec:rsp-construction}
We now present a two-message EB-RSP protocol that uses a free OWGA associated with a certain group. The interesting aspect of this EB-RSP construction is that it does not rely on any trapdoor that would give the verifier a complete classical description of the state held by the prover. Instead, we exploit the algebraic properties of the group action to cancel out parameters unknown to the verifier, ensuring that only parameters known to the verifier appear in the quantum state while still maintaining security. This approach may be of independent interest, as the techniques used here could enable the blind preparation of more specific quantum states with stronger security properties without relying on a trapdoor primitive (as most RSPs rely on variants of TCFs).

We will first describe our two-message EB-RSP construction in \cref{algo:rsp}. Afterwards, we show the correctness and the eavesdropper blindness of the protocol.

{
\floatname{algorithm}{Protocol}
\begin{algorithm}[H]
Let
\begin{itemize}
    \item $\lambda$ be the security parameter
    
    \item $\star : G \times X \to X$ be a free OWGA
    
    \item $G = \Z_p^\lambda$ for a prime $p$, which is poly-sized in the security parameter $\lambda$
\end{itemize}

Protocol Steps:
\begin{enumerate}
    \item $V$ samples $x_0 \leftarrow X$ and $r, s \leftarrow G$. $V$ computes $x_1 := (-s) \star x_0$, and sends $(x_0, x_1, r)$ to $P$.
    
    \item $P$ proceeds as follows:
    \begin{enumerate}
        \item Prepare the state $\sum_{b \in \bits, g \in G} \ket{b, g}$.
        
        \item Evaluate in superposition the function $f(b, g) = g \star x_b$, resulting in the state
        \[
        \sum_{b \in \bits, g \in G} \ket{b, g, g \star x_b}.
        \]
        
        \item Measure the last register in the standard basis, resulting in outcome $y$ and the state
        \[
        \ket{0, g_0} + \ket{1, g_0 + s}.
        \]
        
        \item Evaluate in superposition the function $h(g) = \langle g, r \rangle$, resulting in the state 
        \[
        \ket{0, g_0, \langle g_0, r \rangle} + \ket{1, g_0 + s, \langle g_0 + s, r \rangle}.
        \]
        
        \item Apply QFT over $\Z_p$ to last register and measure to obtain a measurement outcome $z$.
        
        \item Apply QFT over $G$ to second register and measure to obtain a measurement outcome $g$.

        \item Send $(z, g)$ to $V$.
    \end{enumerate}
        
    \item $V$ outputs $\theta := \langle s, z \cdot r + g \rangle \cdot \frac{2\pi}{p}$, and $P$ outputs the post-measurement state $\ket{+_\theta}$.
\end{enumerate}
\caption{Two-Message EB-RSP}
\label{algo:rsp}
\end{algorithm}
}

The main result of this section shows that \cref{algo:rsp} is a two-message EB-RSP (as defined in \cref{def:eb-rsp}):

\begin{ptheorem}\label[ptheorem]{thm:owga-rsp}
    The existence of a free OWGA $\star : G \times X \to X$, where $G = \Z_p^\lambda$ for a prime $p$, which is poly-sized in the security parameter $\lambda$, implies the existence of a two-message EB-RSP with $N = p$.
\end{ptheorem}

\begin{proof}
We will break the proof into two parts: we will first show that \cref{algo:rsp} satisfies correctness in \cref{thm:rsp-correctness} and then show it achieves security, i.e. eavesdropper blindness, in \cref{thm:rsp-blindness}.
\end{proof}

\begin{ptheorem}\label[ptheorem]{thm:rsp-correctness}
   The two-message EB-RSP in \cref{algo:rsp} satisfies correctness.
\end{ptheorem}

\begin{proof}
We will now describe carefully, step by step, the interaction between the classical verifier $V$ and the quantum prover $P$ as well as how the state of $P$ evolves in the protocol. For simplicity, we drop normalization factors in quantum states.
\begin{itemize}
    \item (Verifier 1\textsuperscript{st} Message)
    Sample $x_0 \leftarrow X$ and $r, s \leftarrow G$. Compute $x_1 := (-s) \star x_0$, and send $(x_0, x_1, r)$ to $P$.
    
    \item (Prover 1\textsuperscript{st} Message)
    Prepare the state
    \[
    \ket{+} \otimes \sum_{g \in G} \ket{g}
    = \sum_{b \in \bits, g \in G} \ket{b, g}.
    \]
    Then apply the isometric mapping that lets $g$ act on $x_0$ and $x_1$, respectively, controlled on the first register, to obtain the state
    \[
    \sum_{b \in \bits, g \in G} \ket{b, g, g \star x_b}.
    \]
    Measure the last register in the standard basis to obtain some $y \in X$, with the residual state being
    \[
    \ket{0, g_0} + \ket{1, g_0 + s},
    \]
    where $g_0 \star x_0 = y$, since the group action is free.
    Apply another isometric mapping that computes the inner product between $r$ and the second register, to obtain the state
    \begin{align*}
        \ket{0, g_0, \langle g_0, r \rangle} + \ket{1, g_0 + s, \langle g_0 + s, r \rangle}
        = \ket{0, g_0, z_0} + \ket{1, g_0 + s, z_1},
    \end{align*}
    where we define $z_0 := \langle g_0, r \rangle \in \Z_p$ and $z_1 := \langle g_0 + s, r \rangle \in \Z_p$. Note that $z_1 = z_0 + \langle s, r \rangle \mod{p}$. 
    Now, apply the QFT over $\Z_p$ to the last register to obtain
    \[
    \sum_{z \in \Z_p} (\chi(z, z_0)\ket{0, g_0} + \chi(z, z_1)\ket{1, g_0 + s}) \ket{z},
    \]
    and measure the last register to obtain some $z \in \Z_p$, with the residual state being
    \begin{align*}
        &\phantom{.....} \chi(z, z_0)\ket{0, g_0} + \chi(z, z_1)\ket{1, g_0 + s} \\
        &= \ket{0, g_0} + \chi(z, \langle s, r \rangle)\ket{1, g_0 + s} \\
        &= \ket{0, g_0} + \omega_p^{z \cdot \langle s, r \rangle} \ket{1, g_0 + s} \\
        &= \ket{0, g_0} + \omega_p^{\langle s, z \cdot r \rangle} \ket{1, g_0 + s},
    \end{align*}
    where $z \cdot r$ means that each component of $r$ is multiplied by $z$. Lastly, apply the QFT over $G$ to the second register to obtain
    \[
    \sum_{g \in G} (\chi(g, g_0)\ket{0} + \omega_p^{\langle s, z \cdot r \rangle} \chi(g, g_0 + s) \ket{1}) \ket{g},
    \]
    and measure again the last register to obtain some $g \in G$, with the residual state being
    \begin{align*}
        &\phantom{.....} \chi(g, g_0)\ket{0} + \omega_p^{\langle s, z \cdot r \rangle} \chi(g, g_0 + s) \ket{1} \\
        &= \chi(g, g_0)\ket{0} + \omega_p^{\langle s, z \cdot r \rangle} \chi(g, g_0)\chi(g, s)\ket{1} \\
        &= \ket{0} + \omega_p^{\langle s, z \cdot r \rangle} \chi(g, s)\ket{1} \\
        &= \ket{0} + \omega_p^{\langle s, z \cdot r \rangle + \langle g, s \rangle}\ket{1} \\
        &= \ket{0} + \omega_p^{\langle s, z \cdot r + g \rangle}\ket{1}.
    \end{align*}
    Send $(z, g)$ to $V$. 
    
    \item (Verifier Output) Compute $k := \langle s, z \cdot r + g \rangle \in \Z_p$ and output $\theta = k \cdot \tfrac{2\pi}{p}.$
\end{itemize}
As a result, this shows that the protocol obviously terminates successfully with probability $1$. Furthermore, upon successful completion, the honest prover $P$ holds the state $\ket{+_{\theta}}$ for an angle $\theta \in \Theta_p$ known to the verifier.
\end{proof}

\begin{ptheorem}\label[ptheorem]{thm:rsp-blindness}
    The two-message EB-RSP in \cref{algo:rsp} satisfies eavesdropper blindness.
\end{ptheorem}

\begin{proof}
By the definition of eavesdropper blindness security, we have to show that the distribution of $\theta$ in \cref{algo:rsp} is computationally indistinguishable from the uniform distribution over $\Theta_N$ for $N = p$, from the eavesdropper's view. In other words, we have to show that $k := \langle s, z \cdot r + g \rangle \in \Z_p$ is computationally indistinguishable from the uniform distribution over $\Z_p$. 

Note that during the quantum computation, $g$ is sampled uniformly at random from $G$. Furthermore, $g$ is independent of the measurement outcome $z$ and $V$'s randomness $r$. This implies that $z \cdot r + g$ is uniformly distributed over $G$.

As a result, we can now invoke the quantum version of the Goldreich-Levin theorem for large fields given in \cref{thm:gl-quantum}. More concretely, we instantiate it with $p = \poly(\lambda)$, $H = \Z_p$, and $\Phi$ computes $\star$ with respect to $x_0$. As a result, this shows that if the advantage $\varepsilon$ of the eavesdropper to distinguish between $k := \langle s, z \cdot r + g \rangle$ and a uniform $u \in \Z_p$ would be non-negligible, then this would imply the existence of an efficient extractor for $s$, contradicting the one-wayness of the group action. Consequently, the advantage of the eavesdropper must be negligible, which concludes the security proof. 
\end{proof}

\smallskip\noindent\textbf{Protocol Simplification.}
As also described in \cref{intro:eb-rsp}, the  current protocol can be further simplified by removing $r$ and the measurement outcome $z$ from the construction altogether, which has the same effect as setting $r = 0$, i.e., the all-zero string. At a high-level, the completeness of the new protocol remains unchanged, as it holds for every value of $r$. On the other hand, for the eavesdropper blindness, observe that, for any fixed verifier randomness $r$ and any outcome of the inner product register measurement, the Fourier outcome $g$ is uniformly distributed over $G$. In particular, $g$ is uniform and independent of the verifier's first message. And now one can use the Goldreich-Levin argument for $\langle s, g\rangle$ as before.

\bigskip\noindent\textbf{On Extending the Range of $N$.} So far, we have shown how to construct such EB-RSPs for $N$ being a prime $p$, which is poly-sized in the security parameter $\lambda$. We can, in fact, extend this to $N$ of the form $N = p_1 \cdots p_m$, where the $p_i$ are pairwise distinct poly-sized primes and $m$ is also poly-sized in the security parameter. The idea of constructing such EB-RSPs for larger $N$ is fairly simple: combine the above EB-RSP for different primes. More concretely, view the presented EB-RSP as a subroutine in which the verifier inputs $p$ and the prover inputs a one-qubit state (in this case, it was $\ket{+}$). View the outcome of this subroutine as including a relative phase on the prover's input qubit, while the verifier obtains a classical description of this relative phase. The basic idea is now to run this in parallel for all primes $p_i$. Thus, if we want to construct this EB-RSP for $N = p_1 \cdots p_m$, the verifier simply sends the group action instances together with the randomness $(x_0, x_1, r)$ for each $p_i$ in parallel to the prover, while $P$ performs the subroutine for each $p_i$ sequentially on the same qubit. More precisely, the prover starts with the input $\ket{+}$ and follows the subroutine for $p_1$ to obtain the new one-qubit state $\ket{+_{\theta_1}}$ for $\theta_1 \in \Theta_{p_1}$. Next, the prover again follows the subroutine for $p_2$ by inputting this newly obtained one-qubit state and ends up with $\ket{+_{(\theta_1 + \theta_2)}}$ for $\theta_2 \in \Theta_{p_2}$. This continues until the subroutine has been executed for all $p_i$. Note that in the end $\theta_1 + \ldots + \theta_m \in \Theta_N$ by simply extending the fractions. That this sum is computationally indistinguishable from uniform in $\Theta_N$ follows immediately from the independence of the subroutines, together with the previous security analysis showing that each $\theta_i \in \Theta_{p_i}$ is computationally indistinguishable from uniform, and, lastly, from the fact that all primes are distinct. 

We note that how to extend this idea to powers of primes remains unclear. A naive approach, replacing the QFT over $\Z_p$ with the QFT over prime powers $\Z_{p^m}$, does not work, since correctness fails: the verifier $V$ cannot recover $\theta$ because the phase depends on the term $z_1 - z_0$, and $V$ must know its value modulo $p^m$. Writing $\delta := \langle s, r \rangle$, one has $z_1 - z_0 = \delta$ or $z_1 - z_0 = \delta - p$, depending on the unknown value of $g_0$. Although $V$ can compute $\delta$, it cannot determine which of the two cases holds, since this depends on $g_0$, about which $V$ has no information. Consequently, $V$ cannot compute $z_1 - z_0 \mod p^m$ in general. The case $m = 1$ is the exception, since both possibilities are congruent to $\delta$ modulo $p$.

\bigskip\noindent\textbf{On Extending to General Abelian One-Way Group Actions.} Finally, we want to emphasize that it remains an open problem to construct such EB-RSPs from general abelian OWGAs. We construct them only for certain groups, since we rely on the Goldreich-Levin theorem for large fields. There are indeed generalizations of the Goldreich-Levin theorem to general abelian groups, but the authors are not familiar with the framework in which those theorems are stated. Thus, it may be plausible that more general Goldreich-Levin theorems could extend this construction to more general groups. Nevertheless, this work is still a step forward toward constructing RSP-type primitives without relying on trapdoor primitives, as we exploit the algebraic properties of the group action to compensate for the absence of a trapdoor. Another important aspect is that this EB-RSP has a useful application in QPKE, as we show in \cref{sec:qpke}. It also remains open whether this type of EB-RSP has other useful applications.

%% file: sec-qpke.tex
\section{Quantum Public-Key Encryption with Classical Public Keys}\label{sec:qpke}
In this section, we present a new \emph{quantum public-key encryption} (QPKE) construction with classical public keys and quantum ciphertexts that is IND-CPA secure. The protocol consists of a black-box construction from the previously defined two-message EB-RSP in \cref{sec:rsp} and can therefore ultimately be instantiated from the specific one-way group actions described in \cref{subsec:rsp-construction} or from plain TCFs.

There are many different types of QPKE, depending on whether the public keys and ciphertexts are classical or quantum \cite{TCC:BGHMSV23,EPRINT:BarMalWal23,EC:HhaMorYam23,C:BarKhu25}. As we said, we focus on the case where the public keys are classical and the ciphertexts are quantum, as this setting is still not well understood with respect to minimal cryptographic assumptions. In terms of near-future usability, this setting is also much more practical than having quantum public keys, since one would need to store an enormous number of quantum public keys, which would also be `consumed' after each encryption, thereby creating a need to `refill' them.

We also want to highlight that, since our EB-RSP includes a wide range of output states together with a relatively weak security definition, we can hope for relatively simple constructions from different cryptographic assumptions, thereby allowing the QPKE to be instantiated from a wider range of assumptions. As we have seen in \cref{subsec:rsp-definition}, existing RSPs are likely naturally adaptable to the two-message EB-RSP notion and can therefore subsequently be used to construct QPKE schemes. This yields a QPKE construction from plain TCFs. To compare this result, in some sense, with existing ones, the authors of \cite{C:BarKhu25} used a stronger notion of TCFs, namely \emph{dual-mode TCFs}, to construct a QPKE, but with the advantage of having classical ciphertexts rather than quantum ones.

\subsection{Definition}\label{subsec:qpke-definition}
We start by providing a formal definition of QPKE with classical public keys and quantum ciphertexts. For this, we follow the definition in \cite[Definition 4.8]{EC:HhaMorYam23}.

As in their case, we will define this QPKE for classical one-bit messages for simplicity. The multi-bit message version can be defined analogously, and a simple parallel repetition suffices to expand the message length. Moreover, we can further extend the message space to \emph{quantum states} via hybrid encryption by relying on the quantum one-time pad as in \cite{C:BroJef15}. That is, we encrypt a quantum message using a quantum one-time pad and then encrypt the key of the quantum one-time pad using our QPKE for classical messages.

\begin{pdefinition}[Quantum Public-Key Encryption]\label[pdefinition]{def:qpke}
    A \emph{quantum public-key encryption (QPKE)} scheme (with single-bit messages) consists of the algorithms $(\keygen, \enc, \dec)$:
    \begin{itemize}
        \item $\keygen(1^\lambda) \rightarrow (\pk, \sk)$. This is a PPT algorithm that takes the security parameter $1^\lambda$ as input and outputs a classical public key $\pk$ and a classical secret key $\sk$.

        \item $\enc(\pk, b) \rightarrow ct$. This is a QPT algorithm that takes a public key $\pk$ and a message $b \in \bits$ as input and outputs a quantum ciphertext $ct$.

        \item $\dec(\sk, ct) \rightarrow b'/\perp$. This is a QPT algorithm that takes a secret key $\sk$ and a ciphertext $ct$ as input and outputs a message $b' \in \bits$ or $\perp$.
    \end{itemize}
    
    We require the scheme to satisfy the following properties:
    \begin{itemize}
        \item \emph{(Correctness)} For any $b \in \bits$, we have
        \[
        \Pr{b' = b : (\pk, \sk) \leftarrow \keygen(1^\lambda), ct \leftarrow \enc(\pk, b), b' \leftarrow \dec(\sk, ct)}
        = 1 - \negl(\lambda).
        \]
        
        \item \emph{(IND-CPA Security)} We say the QPKE scheme is \emph{IND-CPA secure} if for any QPT adversary $\Ac$, we have
        \[
        \abs{\Pr{b' = b : b' \leftarrow \Ac(\pk, ct_b)} - \dfrac{1}{2}}
        = \negl(\lambda),
        \]
        where $(\pk, \sk) \leftarrow \keygen(1^\lambda)$, $b \leftarrow \bits$, and $ct_b \leftarrow \enc(\pk, b)$.
    \end{itemize}
\end{pdefinition}

Note that the security definition is given in \emph{guess style}, where the game picks a random bit $b$ and the adversary has to output a guess $b'$ for the bit. Since it aligns more with our eavesdropper blindness security in \cref{def:eb-rsp} that we want to reduce to, we use this formulation.

\subsection{Black-Box Construction from Eavesdropper-Blind Remote State Preparation}\label{subsec:qpke-construction}
We now present the QPKE scheme with classical public keys, built in a black-box fashion based on our two-message EB-RSP defined in \cref{def:eb-rsp}. Before going into the construction, we first prove a lemma that will later help us compute the probability for correctness.

\begin{plemma}\label[plemma]{lem:braket-plus-plus}
    For any $\theta_1, \theta_2 \in \R$, we have
    \[
    \abs{\braket{+_{\theta_2}}{+_{\theta_1}}}^2
    = \tfrac{1}{2} + \tfrac{\cos(\theta_1 - \theta_2)}{2}.
    \]
\end{plemma}

\begin{proof}
We calculate:
\begin{align*}
    \braket{+_{\theta_2}}{+_{\theta_1}}
    &= \tfrac{1}{2} \parens{\bra{0} + e^{-i\theta_2}\bra{1}} \parens{\ket{0} + e^{i\theta_1}\ket{1}} \\
    &= \tfrac{1}{2} \parens{1 + e^{i\parens{\theta_1 - \theta_2}}}.
\end{align*}
Hence,
\begin{align*}
    \abs{\braket{+_{\theta_2}}{+_{\theta_1}}}^2
    &= \braket{+_{\theta_2}}{+_{\theta_1}} \cdot \braket{+_{\theta_1}}{+_{\theta_2}} \\
    &= \tfrac{1}{4} \parens{1 + e^{i\parens{\theta_1 - \theta_2}}} \parens{1 + e^{i\parens{\theta_2 - \theta_1}}} \\
    &= \tfrac{1}{4} \parens{1 + e^{i\parens{\theta_2 - \theta_1}} + e^{i\parens{\theta_1 - \theta_2}} + 1} \\
    &= \tfrac{1}{2} + \tfrac{\cos(\theta_1 - \theta_2)}{2}.
\end{align*}
\end{proof}

The main result of this section is stated in the following theorem.

\begin{ptheorem}\label[ptheorem]{thm:qpke}
    The existence of a two-message EB-RSP implies the existence of an IND-CPA secure QPKE scheme.
\end{ptheorem}

\smallskip\noindent\textbf{QPKE Scheme.} We first define 
\[
n 
= \begin{cases}
1, & \text{if } 2 \mid N, \\
2\lambda + 1, & \text{if } 2 \nmid N.
\end{cases}
\]
The key generation, encryption, and decryption algorithms are given by:
\begin{itemize}
    \item $\keygen(1^\lambda)$: Run $\ebrsp.\setup(1^\lambda)$ exactly $n$ times and output $\pk := (\msg_{V_1}, \ldots, \msg_{V_n})$ and $\sk := (\st_{V_1}, \ldots, \st_{V_n})$.

    \item $\enc(\pk, b)$: Given a message bit $b \in \bits$, run $\ebrsp.\qcomp(\msg_{V_i})$ for each $i$ to obtain $(\ket{+_{\theta_1}}, \ldots, \ket{+_{\theta_n}}, \msg_{P_1}, \ldots, \msg_{P_n})$. Now apply $U := R_z(b \cdot \floor{\tfrac{N}{2}} \cdot \tfrac{2\pi}{N})$ to all quantum states and output
    \[
    ct
    := (U \ket{+_{\theta_1}}, \ldots, U \ket{+_{\theta_n}}, \msg_{P_1}, \ldots, \msg_{P_n}).
    \]

    \item $\dec(\sk, ct)$: Recover the $\theta_i$'s by running $\ebrsp.\dec(\st_{V_i}, \msg_{P_i})$ for each instance. Apply the unitary $U_i := R_z(-\theta_i)$ to the $i$-th qubit, which are now all in the state $U \ket{+}$. Measure all of them in the Hadamard basis, yielding outcomes $0$ for $\ket{+}$ and $1$ for $\ket{-}$. Finally, output the bit that appears most often using a majority vote.
\end{itemize}

\begin{ptheorem}
    The QPKE scheme as described above is correct.
\end{ptheorem}

\begin{proof}
We perform a case distinction depending on the parity of $N$.

We start with the case where $N$ is even. Note that in this case we have $\floor{\tfrac{N}{2}} \cdot \tfrac{2\pi}{N} = \pi$. Hence $U = R_z(b\pi)$. Therefore the received qubit is in the state $\ket{+_{(\theta_1 + b\pi)}}$. In the decryption algorithm we remove $\theta_1$, as $R_z(-\theta_1)$ is applied to the qubit. We are left with the state $\ket{+_{b\pi}} = H \ket{b}$. Measuring this in the Hadamard basis then yields $b$ with probability $1$.

Now consider the case where $N$ is odd and the encrypted message was $b = 0$. In this case $U = I$ and all states in the decryption phase right before the measurement are in the $\ket{+}$ state, which all output $0$ under a Hadamard measurement. By majority vote, $0$ will be the output of the decryption. Hence the decryption again succeeds with probability $1$.

Lastly, consider the case where $N$ is odd and the encrypted message was $b = 1$. This case is the reason we included repetition within the encryption algorithm in order to boost the correctness probability. We observe that
\[
b \cdot \floor{\tfrac{N}{2}} \cdot \tfrac{2\pi}{N} 
= \pi - \tfrac{\pi}{N} 
=: \theta'.
\]
Hence all qubits right before the Hadamard measurement are in the state $\ket{+_{\theta'}}$. We first compute the probability that one of the measurements yields the correct bit. Our measurement outcome must correspond to $\ket{-}$, since $b = 1$ was encrypted. The probability that we measure $\ket{-}$ is given by
\begin{align*}
    \abs{\braket{-}{+_{\theta'}}}^2
    &= \abs{\braket{+_\pi}{+_{\theta'}}}^2 \\
    &= \tfrac{1}{2} + \tfrac{\cos(\theta' - \pi)}{2} \\
    &= \tfrac{1}{2} + \tfrac{\cos(\tfrac{\pi}{N})}{2} \\
    &\geq \tfrac{1}{2} + \tfrac{\cos(\tfrac{\pi}{3})}{2} \\
    &= 0.75, 
\end{align*}
where we used \cref{lem:braket-plus-plus} for the second equality and lower-bounded the expression since $N$ is odd and, without loss of generality, at least $3$, and the cosine function is strictly monotonically decreasing on $[0, \tfrac{\pi}{2}]$. Since the probability that a single outcome is correct is at least $75\%$, a standard Chernoff bound argument implies that the probability that the majority vote decrypts correctly is $1 - \negl(\lambda)$.
\end{proof}

\begin{ptheorem}
    The QPKE scheme as described above is IND-CPA secure.
\end{ptheorem}

\begin{proof}
We reduce the security of the QPKE scheme to the security of the EB-RSP. An adversary $\Ac$ against the QPKE security game receives the public key $\pk := (\msg_{V_1}, \ldots, \msg_{V_n})$ and the ciphertext $ct_b$, which encrypts a bit $b \in \bits$. Instead of giving the adversary only the quantum states, we even provide $\Ac$ with the classical description of the quantum states given by
\[
\theta_i + b \cdot \floor{\tfrac{N}{2}} \cdot \tfrac{2\pi}{N}
= \parens{k_i + b \cdot \floor{\tfrac{N}{2}}} \cdot \tfrac{2\pi}{N},
\]
where $\theta_i = k_i \cdot \tfrac{2\pi}{N}$ with $k_i \in \Z_N$. This only increases the advantage of $\Ac$, and we show that it still remains negligible. Thus the adversary receives
\[
\parens{\msg_{V_i}, \msg_{P_i}, k_i + b \cdot \floor{\tfrac{N}{2}} \mod{N}}_{1 \leq i \leq n}.
\]
Note that the first two elements correspond exactly to the transcript of the two-message EB-RSP. Since the EB-RSP is eavesdropper-blind, we know that $k_i$ is computationally indistinguishable from a uniform element of $\Z_N$. By a hybrid argument, the information above is computationally indistinguishable from
\[
\parens{\msg_{V_i}, \msg_{P_i}, u_i}_{1 \leq i \leq n},
\]
where $u_i \leftarrow \Z_N$, as all instances are independent of each other. Hence the information about $b$ is completely hidden and the adversary has no non-negligible advantage.
\end{proof}

This concludes the proof that two-message EB-RSP implies IND-CPA secure QPKE with classical public keys. Combining this theorem with our construction of two-message EB-RSP from specific OWGAs yields the following corollary.

\begin{pcorollary}
    The existence of a free OWGA $\star : G \times X \to X$, where $G = \Z_p^\lambda$ for a prime $p$, which is poly-sized in the security parameter $\lambda$, implies the existence of an IND-CPA secure QPKE scheme.
\end{pcorollary}

\begin{proof}
This follows immediately from \cref{thm:owga-rsp,thm:qpke}.
\end{proof}

Furthermore, the relationship between two-message EB-RSP and IND-CPA secure QPKE yields additional constructions of QPKE from another well-studied assumption:

\begin{pcorollary}
    The existence of a plain TCF implies the existence of an IND-CPA secure QPKE scheme.
\end{pcorollary}

\begin{proof}
This follows immediately from \cref{thm:tcf-rsp,thm:qpke}.
\end{proof}

\smallskip\noindent\textbf{On the Security of the EB-RSP for QPKE.} Our EB-RSP security definition considers the full angle $\theta = k \cdot \tfrac{2\pi}{N}$ with $k \in \Z_N$ to be computationally indistinguishable from uniform. When $N$ is a power of $2$, this means that each individual bit of $k$ should be computationally indistinguishable from uniform. 

But note that for IND-CPA security, we only need the most significant bit of $k$ to be computationally indistinguishable, since our encryption essentially flips the most significant bit of $k$ within the quantum state when $N$ is a power of $2$. For general $N$, this translates to effectively hiding the single bit of information corresponding to whether $0 \leq k < \floor{\tfrac{N}{2}}$ or $\floor{\tfrac{N}{2}} \leq k < N$, that is, whether $\theta$ lies on the upper half or the lower half of the unit circle. Thus, requiring only that this information be hidden would be sufficient to ensure IND-CPA security. This is exactly equivalent to hiding the most significant bit of $k$ when $N$ is a power of $2$. An interesting fact is that all other known RSPs do it the other way around: they try to hide all but the most significant bit (since it is impossible to hide everything while allowing the prover itself to be malicious).

%% file: sec-appendix.tex
\section{Eavesdropper-Blind Remote State Preparation from Trapdoor Claw-Free Functions}\label{sec:appendix}
In this section, we provide a description of the RSP protocol presented in \cite{C:BKMSW25}, which relies solely on the existence of a plain trapdoor claw-free function. As described in the paragraph before \cref{thm:tcf-rsp}, this RSP can be modified to obtain a two-message EB-RSP from plain trapdoor claw-free functions.

Their RSP constructs states of the form
\[
\ket{+_\theta} 
:= \dfrac{1}{\sqrt{2}}(\ket{0} + e^{i \theta}\ket{1}),
\]
where $\theta \in \Theta := \{k \cdot \pi/4 \mid k = 0, \ldots, 7\}$, and achieves the standard blind security notion against any malicious quantum prover. More precisely, they construct states of the form $Z^b \ket{+_\theta} = \ket{+_{(\theta + b\pi)}}$, while the security guarantee is provided only for $\theta$.

We first recall the definition of a trapdoor claw-free function.
\begin{pdefinition}[Trapdoor Claw-Free Function]
    Let $\lambda$ be the security parameter. A \emph{trapdoor claw-free function (TCF)} consists of a family of injective function pairs $(f_{0,\lambda}, f_{1, \lambda})$ and finite sets $\mathcal{X}_\lambda$ and $\mathcal{Y}_\lambda$ with
    \[
    \left\{f_{b, \lambda} : \mathcal{X}_\lambda \to \mathcal{Y}_\lambda \right\}_{(b, \lambda) \in \bits \times \mathbb{N}},
    \]
    where we omit the subscript $\lambda$ when it is clear from the context. Additionally, a TCF pair is augmented with two algorithms.
    \begin{itemize}
        \item $\Gen(1^\lambda)$: On input the security parameter in unary $1^\lambda$, the polynomial-time generation algorithm outputs a function pair $(f_0, f_1)$ and a trapdoor $\td$.
        \item $\Invert(\td, y)$: On input an image $y\in \mathcal{Y}$ and the trapdoor $\td$, the polynomial-time deterministic inversion algorithm returns two preimages $(x_0, x_1)$.
    \end{itemize}
     We require a TCF to satisfy the following properties:
     \begin{itemize}
        \item \emph{(Correctness)} For all $\lambda\in \mathbb{N}$, all $x \in \mathcal{X}$, and all $b\in\{0,1\}$, it holds that:
        \[
        f_0(x_0) = f_1(x_1) = y,
        \]
        where $((f_0, f_1), \td)\gets \Gen(1^\lambda)$ and $(x_0, x_1)\gets\Invert(\td, f_b(x))$.

        \item \emph{(Efficient Superposition)} There exists a QPT algorithm that, on input the description of the functions $(f_0, f_1)$, prepares the state
        \[
        \frac{1}{\sqrt{|\mathcal{X}|}}\sum_{x\in \mathcal{X}} \ket{x}.
        \]

        \item \emph{(Claw-Freeness)} For all QPT algorithms $A^*$ there exists a negligible function $\negl$ such that for all $\lambda \in \mathbb{N}$ it holds that:
        \[
        \Pr{(x_0^*, x_1^*) \gets A^*(f_0, f_1): f_0(x_0^*) = f_1(x_1^*)} \leq \negl(\lambda).
        \]
        where $((f_0, f_1), \td)\gets \Gen(1^\lambda)$.
     \end{itemize}
     
     We also assume that there exists an embedding of the set $\mathcal{X}$ into the bitstrings $\{0,1\}^{p(\lambda)}$ for some fixed polynomial $p$.
\end{pdefinition}

Next, we present their main RSP protocol, assuming the existence of any TCF $(\Gen, \Invert)$. They start with describing a subroutine that the prover $P$ and the verifier $V$ will run in their main protocol.

\bigskip\noindent\textbf{Subroutine.} The input and output of the protocol are:
\begin{itemize}
    \item (Input) The protocol is parameterized by the security parameter in unary $1^\lambda$, an integer $n$ and the prover $P$ holds a state $\ket{\psi} = \alpha \ket{0} + \beta \ket{1}$.
    
    \item (Output) At the end of the interaction, the verifier holds a pair $(b, \theta) \in \{0,1\} \times \{0,1,2\}$ and the prover holds the state $\alpha\ket{0} + \beta (-1)^b \omega_n^\theta\ket{1}$.
\end{itemize}

The interaction between $P$ and $V$ proceeds as follows:
\begin{itemize}
    \item (Verifier 1\textsuperscript{st} Message) Sample $((f_0, f_1), \td) \gets \Gen(1^\lambda)$ and send $(f_0, f_1)$ to $P$.
    
    \item (Prover 1\textsuperscript{st} Message) Prepare the state
    \[
    \ket{\psi} \otimes \frac{1}{\sqrt{|\mathcal{X}|}} \sum_{x \in \mathcal{X}} \ket{x} = 
    \frac{1}{\sqrt{|\mathcal{X}|}} \sum_{x \in \mathcal{X}} \alpha \ket{0, x} + \beta \ket{1, x}.
    \]
    Then, apply the isometric mapping that evaluates $f_b$ coherently on input the second register, with the function controlled on the first register, to obtain the state
    \[
    \frac{1}{\sqrt{|\mathcal{X}|}} \sum_{x \in \mathcal{X}} \alpha \ket{0, x, f_0(x)} + \beta \ket{1, x, f_1(x)}.
    \]
    Measure the last register to obtain some $y \in \mathcal{Y}$, with the residual state being
    \[
    \alpha \ket{0, x_0} + \beta \ket{1, x_1}
    \]
    where $f_0(x_0) = f_1(x_1) = y$. Send $y$ to $V$.
    
    \item (Verifier 2\textsuperscript{nd} Message) Check if $y \in \mathcal{Y}$ and abort if not. Sample two strings $r_0, r_1 \sample \{0,1\}^{p(\lambda)}$ uniformly at random. Send $(r_0, r_1)$ to $P$.
    
    \item (Prover 2\textsuperscript{nd} Message) Consider the isometric mapping 
    \[
    M:(b, x_b) \mapsto (b, x_b, (-1)^{1-b} (x_b \cdot r_b))
    \]
    where the inner product $z_b := x_b \cdot r_b \in \bits$ is computed over $\Z_2$ and then parsed as an element of $\Z_n$. Apply $M$ to the current state to compute
    \[
    \alpha \ket{0, x_0, - (x_0 \cdot r_0)} + \beta \ket{1, x_1, x_1 \cdot r_1} =
    \alpha \ket{0, x_0, - z_0} + \beta \ket{1, x_1, z_1}.
    \]
    Apply QFT$_n$ to the last register to obtain
    \[
    \frac{1}{\sqrt{n}} \sum_{d'\in \Z_n} (\omega_n^{-d'\cdot z_0}\alpha\ket{0, x_0} + \omega_n^{d' \cdot z_1} \beta\ket{1, x_1}) \ket{d'},
    \]
    where $-d' \cdot z_0, d' \cdot z_1 \in \Z_n$. Measure the last register in the computational basis and abort if the output $d'\neq 1$. The state becomes
    \[
    \omega_n^{-z_0}\alpha\ket{0, x_0} + \omega_n^{z_1} \beta\ket{1, x_1}
    \equiv \alpha\ket{0, x_0} + \omega_n^{z_0 + z_1} \beta\ket{1, x_1}.
    \]
    Conditioning on not aborting, measure the second register in the Hadamard basis to obtain some $d\in\{0,1\}^{p(\lambda)}$, and return the state
    \[
    \alpha \ket{0} + \beta (-1)^{d \cdot (x_0 \oplus x_1)} \omega_n^{z_0 + z_1} \ket{1}.
    \]
    Send $d$ to $V$.
    
    \item (Verifier Output) Recompute $(x_0, x_1) \gets \Invert(\td, y)$ and set $b := d \cdot (x_0 \oplus x_1)$ and $\theta := z_0 + z_1 = x_0 \cdot r_0 + x_1 \cdot r_1 \in \{0,1,2\}$, where the sum is computed over $\Z$.
\end{itemize}

\bigskip\noindent\textbf{Main Protocol.} The main protocol uses the above defined subroutine and proceeds in three steps.
\begin{itemize}
    \item Run the above protocol with $n = 2$ and set $\ket{+}$ to be $P$'s input state. Let $(b_1, \theta_1)$ be the output of $V$, and let $\ket{\psi_1}$ be the output of $P$.
    
    \item Run the above protocol with $n = 4$ and set $\ket{\psi_1}$ to be $P$'s input state. Let $(b_2, \theta_2)$ be the output of $V$, and let $\ket{\psi_2}$ be the output of $P$.
    
    \item Run the above protocol with $n = 8$ and set $\ket{\psi_2}$ to be $P$'s input state. Let $(b_3, \theta_3)$ be the output of $V$, and let $\ket{\psi_3}$ be the output of $P$.
\end{itemize}
The prover $P$ returns the final state $\ket{\psi_3}$, whereas the verifier $V$ sets 
\[
b := b_1 \oplus b_2 \oplus b_3 \text{ and }\theta := 4\theta_1 + 2\theta_2 + \theta_3 \mod{8}
\]
and must multiply $\theta$ by $\pi/4$ to obtain the angle.

\bigskip
For a full analysis of the correctness and security of their protocol, we refer the reader to their paper \cite{C:BKMSW25}.